\documentclass[twocolumn,aps,prl,showpacs]{revtex4}
\usepackage[dvips]{graphicx}
\begin{document}
\title{Heisenberg Dimer  Single Molecule Magnets in a Strong Magnetic Field}
\author{Dmitri V. Efremov}
\email{efremov@mpipks-dresden.mpg.de}
\author{Richard A. Klemm}
\email{rklemm@mpipks-dresden.mpg.de}
\affiliation{Max-Planck-Institut f{\"u}r Physik komplexer Systeme,
N{\"o}thnitzer
 Stra{\ss}e 38, D-01187 Dresden, Germany}
\date{\today}
\begin{abstract}
We calculate the static and dynamic properties  of  single
crystal, 
single molecule magnets 
consisting of equal spin $S=1/2$ or 5/2 dimers.  The spins in each dimer interact with
each other
 via the Heisenberg exchange interaction and with the magnetic induction ${\bf B}$ via the
Zeeman interaction, and interdimer interactions are negligible.   For
antiferromagnetic couplings, 
 the static magnetization and
specific heat  exhibit interesting
low temperature $T$  and strong ${\bf B}$ quantum effects.     We calculate 
the frequency
spectrum of the Fourier transform of the real part of the time autocorrelation
function ${\cal C}_{11}(t)$ for arbitrary $T$, ${\bf B}$, and compare our 
results  with those
obtained for classical spins. We also calculate the inelastic neutron magnetic
dynamical structure factor $S({\bf q},\omega)$ at arbitrary $T$, ${\bf B}$.
\end{abstract}
\pacs{05.20.-y, 75.10.Hk, 75.75.+a, 05.45.-a}
\vskip0pt\vskip0pt
\maketitle

\section{Introduction}

Recently, there has been a considerable interest  in the physics of single
molecule magnets (SMM's), or  magnetic
molecules.\cite{Mn12,Fe8,Fe6,Fe10,V2,V2neutron,V2P2O9,Cu2,Yb2,Cr2,Fe2,Fe2structure}  These consist
of small clusters of
 magnetic ions embedded within a
non-magnetic ligand group, which may crystallize into large, well-ordered
single crystals of sufficient quality for neutron scattering
studies.   Usually, the spins within a single molecule interact mainly via the Heisenberg exchange
interaction.  In the simplest SMM's, V2, Cu2, two examples of
Yb2, Cr2,  and four
examples of Fe2, the magnetic cores of the molecules
 consist of dimers of spin $S=1/2$ V$^{4+}$, Cu$^{2+}$, or Yb$^{3+}$, $S=3/2$ Cr$^{3+}$, or $S=5/2$ Fe$^{3+}$ spins,
respectively.\cite{V2,Cu2,Yb2,Cr2,Fe2}  Low-field magnetization, nuclear magnetic
resonance (NMR) and electron paramagnetic resonance (EPR) experiments, and
zero-field inelastic neutron scattering experiments were made on some of these
dimers. \cite{V2,V2neutron,V2P2O9,Cu2,Yb2,Cr2,Fe2,Fe2structure}
Theoretically, zero-field results for the time dependence of the
autocorrelation function of the quantum $S=1/2$ and $S=5/2$
dimers were presented.\cite{MSL}  

Here we study the simplest model of interacting Heisenberg spins in a
magnetic field ${\bf H}$.  We assume only two spins, which interact with each other via
the ordinary Heisenberg exchange interaction $J$, and also with a constant
magnetic induction ${\bf B}$ induced by the application of ${\bf H}$.  For
simplicity, we limit our discussion to the spin values
$S=1/2$ and $S=5/2$. 

We first evaluate the static magnetization ${\bf M}$ and specific heat $C_V$
as  functions
 of $T$ and
${\bf B}$.  
We find that for ferromagnetic (FM) exchange couplings $J>0$, both ${\bf M}$ and
$C_V$ behave at low $T$ as do a single paramagnetic ion with spin $2S$,
qualitatively similar to that expected from a classical treatment.  For
antiferromagnetic (AFM) exchange couplings $J<0$, however, for  $k_BT\ll|J|$, the low-$T$
results  are very non-classical, even for $S=5/2$.  ${\bf M}({\bf B})$ for the
AFM 
 spin-$S$ dimer exhibits
$2S$ discrete steps, reminiscent of the transverse conductivity in the
integer quantum Hall effect.  In addition,  $C_V(T,B)$ 
exhibits $2S$ doublet peaks centered about the corresponding magnetization
step fields,
 the splitting of which is
proportional to $T$.

At arbitrary $T, {\bf B}$, we then evaluate the real part of the time autocorrelation function ${\cal
C}_{11}(t)$, and focus upon its Fourier transform $\tilde{\cal
C}_{11}(\omega)$, which is applicable to inelastic neutron
scattering experiments.  For ${\bf B}=0$, $\tilde{\cal
C}_{11}(\omega)$ for  the spin-$S$ dimer has $4S+1$
equally-spaced modes with frequencies $\omega^0_{S,n}=n|J|$, where $n=-2S,-2S+1,...,2S-1,2S$.
Depending upon $T$ and the sign of $J$, the relative importance of these modes
varies significantly. 
 For ${\bf B}\ne0$, each of these  $\omega^0_{S,n}$ modes is split into 3 modes.
We study the $T$ and ${\bf B}$ dependence of the most important of these modes
for $S=1/2$ and $5/2$, for both FM and AFM exchange couplings. 
For the FM case, only a few modes are important at low $T$, and their relative
strength is nearly independent of
$|B|$.  For the AFM case, however, the situation is  more
complicated, as many modes can be important at rather low $T$, and their
relative importance shifts with $|B|$.  For comparison, we also present the analogous results
for classical spins.   Finally,  at arbitrary $T, {\bf B}$, we evaluate the magnetic dynamical structure factor
$S({\bf q},\omega)$ measurable with inelastic neutron scattering, and identify a method by which it
can measure  $\tilde{\cal C}_{11}(\omega)$.

\section{Thermodynamic Properties}

\subsection{Partition Function}

Here we derive the partition function $Z$ for the quantum dimer for ${\bf
B}\ne0$ , with 
spin $i=1,2$ represented by the operator ${\bf S}_i$, and
set $\hbar=1$ for convenience.
The Hamiltonian is
\begin{eqnarray}
{\cal H}&=&-J{\bf S}_1\cdot{\bf S}_2-\gamma {\bf B}\cdot({\bf S}_1+{\bf S}_2),
\end{eqnarray}
where $\gamma=g\mu_B$ is the gyromagnetic ratio.
Letting
the total spin operator ${\bf s}={\bf S}_1+{\bf S}_2$, ${\cal H}$ is rewritten as
\begin{eqnarray}
{\cal H}&=&-J({\bf s}^2-{\bf S}_1^2-{\bf S}_2^2)/2-\gamma Bs_z.
\end{eqnarray} 
The
dimer quantum states are then indexed by the quantum numbers $s$ and $m$, where
\begin{eqnarray}
{\bf s}^2|sm\rangle&=&s(s+1)|sm\rangle,\\
s_z|sm\rangle&=&m|sm\rangle,\label{m}
\end{eqnarray}
and since 
${\bf S}_i^2|sm\rangle=S(S+1)|sm\rangle$, where $S(S+1)$ is
a constant for all measurable quantities, we drop the terms proportional to
${\bf S}_1^2$ and ${\bf S}_2^2$ in ${\cal H}$ for convenience.

\begin{figure}[t]
\includegraphics[width=0.45\textwidth]{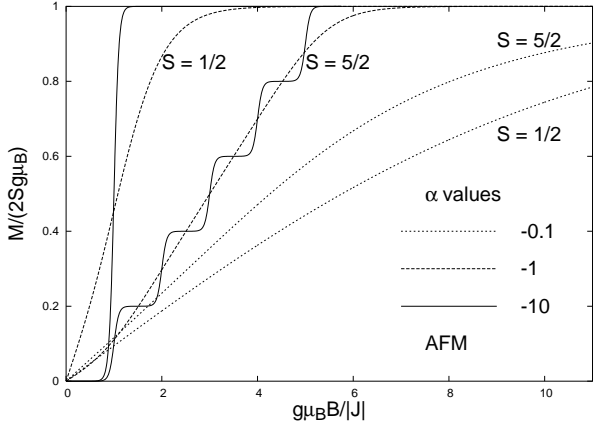}
\caption{Plots of the static magnetization $M$ normalized to $2Sg\mu_B$ versus $\overline{B}=g\mu_BB/|J|$
for the AFM $S=1/2$ and $S=5/2$ dimers at
$\alpha=J/(2k_BT)=-0.1, -1, -10$, as indicated.}
\label{fig1} 
\end{figure}

Letting $\beta=1/(k_BT)$, $\alpha=\beta J/2$, $b=\beta\gamma B$, and
$\overline{B}=g\mu_BB/|J|=b/(2|\alpha|)$, the partition function $Z={\rm
Tr}\exp(-\beta{\cal H})$ for the Heisenberg dimer  is simply
\begin{eqnarray}
Z&=&\sum_{s=0}^{2S}\sum_{m=-s}^se^{\alpha s(s+1)+bm}=\sum_{s=0}^{2S}{\cal
D}_s(\alpha,b),\label{pf}\\
\noalign{\rm where}\nonumber\\
{\cal D}_S(\alpha,b)&=&e^{\alpha
s(s+1)}{{\sinh[(2s+1)b/2]}\over{\sinh(b/2)}}.\label{dos}
\end{eqnarray}

\subsection{Magnetization}

The magnetization ${\bf M}=\beta^{-1}\overrightarrow{\nabla}_{\bf
B}(\ln{Z})$ for the spin-$S$ dimer in a field 
is then easily found to be
\begin{eqnarray}
{\bf M}(\alpha,b)&=&{{\gamma\hat{\bf B}}\over{Z}}\sum_{s=0}^{2S}{\cal D}_s(\alpha,b)B_s(b)\label{magnetization}\\
&\equiv&
\gamma\hat{\bf B}\langle B_s(b)\rangle,\label{average}\\
\noalign{\rm where}\nonumber\\
B_s(b)&=&(s+1/2)\coth[(s+1/2)b]-{1\over{2}}\coth(b/2)\nonumber\label{bsofb}\\
\end{eqnarray}
is the standard Brillouin function for a spin $s$ paramagnet in a
 magnetic field. 
We note that
$\langle\ldots\rangle$ represents
a thermodynamic average over the dimer quantum  $s$ values, evaluated using the weighting function
${\cal D}_s(\alpha,b)$.

In Fig. 1, we plotted $M/(2Sg\mu_B)$ as a function of
$\overline{B}$ for the AFM $S=5/2$ and $S=1/2$ dimers at 
$\alpha=-0.1, -1, -10$, respectively.  At high $k_BT/|J|$
($|\alpha|<<1$), the dotted curves for $\alpha=-0.1$, $S=1/2$, 5/2 are smooth functions of
$\overline{B}$, not too different from the analogous results obtained for classical spins.  At $\alpha=-1$, the dashed curves remain smooth, but
are shifted
over to smaller field values, with hints of a kink just below the saturation
magnetization value.  Most interesting are the low-$T$ effects.  At $\alpha=-10$, 
$M$ has $2S$ steps at integral values
of $\overline{B}$, as shown by the solid curves for $S=1/2,5/2$  in Fig. 1.
Since these steps are thermodynamic, they are reversible.
However, a measurement  at fixed $B$ near to a step value can lead to interesting, non-monotonic
behavior of $M(T)$ at low $T$.  

These thermodynamic steps in $M$ are a consequence
of quantum level crossing due to the strong $B$.\cite{Fe10}
The energy $E_{sm}$ of the state $|sm\rangle$ is given by 
\begin{eqnarray}
E_{sm}&=&-Js(s+1)-m\tilde{B}.
\end{eqnarray}
where $\tilde{B}=\gamma B$ and $\overline{B}=\tilde{B}/|J|$.
For the AFM case, $J=-|J|$, the lowest energy state for each $s$ value is
$E_{sm,\rm min}=E_{ss}$.
The difference in energy between the lowest energy state with quantum number
$s$ and that with the next highest quantum number, $s+1$, is then
\begin{eqnarray}
\Delta E_{ss}&=E_{s+1,s+1}-E_{ss}=|J|(s+1-|\overline{B}|),\label{deltaE}
\end{eqnarray}
which vanishes at $|\overline{B}|=s+1$.  For $S=1/2$, there will only be one
step, as this crossing can only occur between states corresponding to the
$s=0$ and $s=1$ quantum numbers.  Similarly, for $S=5/2$, there will be 5
level crossings, corresponding to $s=0,\ldots,4$.

For the FM case, the situation is rather boring by comparison.  We found that
for both $S=1/2, 5/2$, $M(\alpha,b)/\gamma$ is closely approximated by $B_{2S}(b)$.  For
$\alpha=0.1$, this approximation is accurate to a few percent, but for
$\alpha=10$, the corresponding  curves for $M$ and $B_{2S}$ are indistinguishable.   This is because the $s=2S$ term in both the
numerator and the denominator of 
$\langle B_s\rangle$  is dominant for $J>0$. 
						   
\begin{figure}[floatfix]
\includegraphics[width=0.45\textwidth]{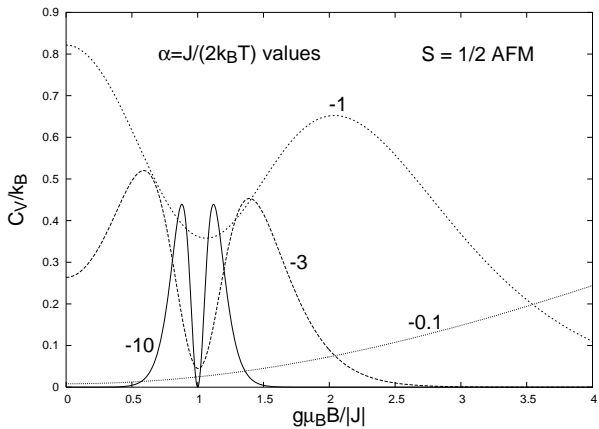}
\caption{Plots of $C_V/k_B$ versus $\overline{B}$ at
$\alpha=-0.1,-1,-3,-10$, as indicated, for the $S=1/2$ AFM dimer.}
\label{fig2} 
\end{figure}

\begin{figure}[t]
\includegraphics[width=0.45\textwidth]{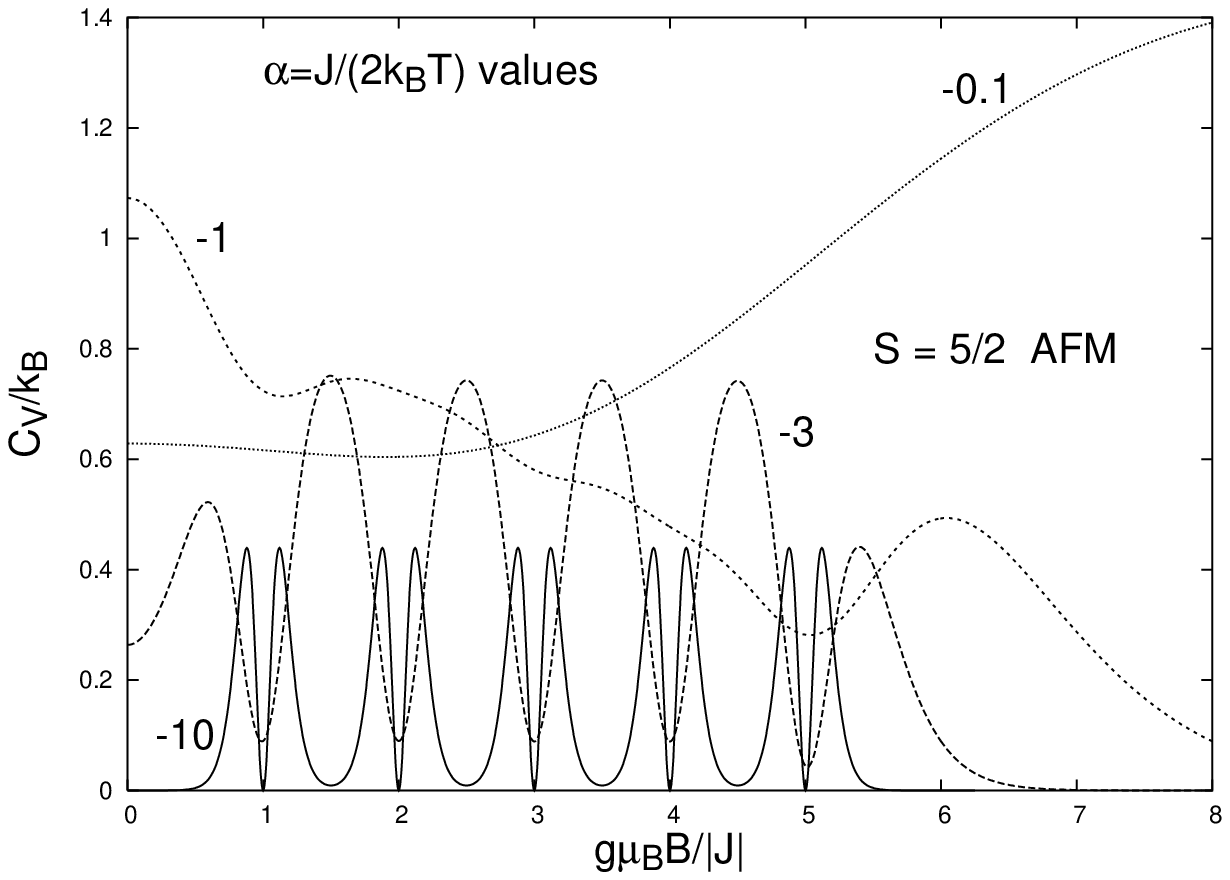}
\caption{Plots of $C_V/k_B$ versus $\overline{B}$ at
$\alpha=-0.1,-1,-3,-10$, as indicated, for the $S=5/2$ AFM dimer.}
\label{fig3} 
\end{figure}

\begin{figure}[b]
\includegraphics[width=0.45\textwidth]{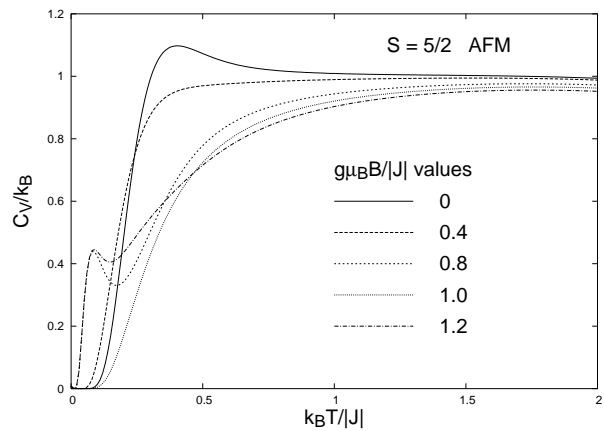}
\caption{Plots of $C_V/k_B$ versus $k_BT/|J|$ at
$\overline{B}=0, 0.4, 0.8, 1.0, 1.2$, as indicated, for the $S=5/2$ AFM dimer. }
\label{fig4} 
\end{figure}

\subsection{Specific Heat}

The specific heat $C_V$ is given by 
\begin{eqnarray}
k_BT^2C_V&=&{{\partial^2\ln{Z}}\over{\partial\beta^2}},
\end{eqnarray}
leading to
\begin{eqnarray}
C_V/k_B&=& b^2\Bigl[\langle s(s+1)\rangle-\coth(b/2)\langle B_s(b)\rangle-\langle
B_s(b)\rangle^2\Bigr]\nonumber\\
& &+2\alpha b\Bigl[\langle s(s+1)B_s(b)\rangle-\langle B_s(b)\rangle\langle
s(s+1)\rangle\Bigr]\nonumber\\
& &+\alpha^2\Bigl[\langle [s(s+1)]^2\rangle-\langle
s(s+1)\rangle^2\Bigr].\label{heatcap}
\end{eqnarray}

In numerical evaluations of $C_V$, we checked that Boltzmann's law,
$\int_0^{\infty}dTC_V/T=2k_B\ln(2S+1)$, is satisfied.
As for the FM $M$, $C_V$ for the FM spin-$S$ dimer is rather boring.
It is closely approximated by the $s=2S$ terms in each of the sums.  In this
approximation, only the first term proportional to $b^2$ survives, so $C_V$
reduces to the standard Schottky result for a spin $2S$ paramagnet,
\begin{eqnarray}
C_V/k_B&{{\rightarrow}\atop{\alpha\gg1}}&C_{2S}(b),\\
&=&b^2\Bigl(2S(2S+1)+{1\over{4}}\coth^2(b/2)\nonumber\\
& &-(2S+1/2)^2\coth^2[(2S+1/2)b]\Bigr).
\end{eqnarray}
Numerically, for $S=5/2$, the FM $C_V$  is nearly indistinguishable
from 
$C_{2S}(b)$ for $\alpha\ge1$.  We note that $C_{2S}(b)$ grows as
$2S(2S+1)b^2/3$ for $b\ll1$, decays as  $b^2\exp(-b)$ for $b\gg1$, and has
a maximum for $b\approx1$.  For $S=5/2$, the maximum
occurs at $b\approx0.78$.

On the other hand, $C_V$ for the AFM spin-$S$ dimer is much more interesting.
In Fig. 2 and 3, we  plotted $C_V/k_B$ versus $\overline{B}$
at $\alpha = -0.1, -1, -3, -10$, for dimers of spin $S=1/2$ and 5/2,
respectively.  The curves for $\alpha=-0.1$ have broad
maxima at fields too high to appear in these figures.  But, as $T$ is lowered 
to $\alpha=-1$, Fig. 2 illustrates that this broad maximum
develops into two peaks centered at $\overline{B}\approx 0,2$, respectively.
Then, as $T$ is lowered further, $C_V$ for $S=1/2$ becomes two
well-defined peaks symmetrically centered about $\overline{B}=1$, with a
splitting between them proportional to $T$. 

In Fig. 3, the analogous AFM $C_V$  for $S=5/2$ is shown.  For
$\alpha=-1$, instead of two broad peaks, as in Fig. 2, for $S=5/2$ there is an irregular
pattern resulting from many accessible energy levels.  The situation becomes clearer
at $\alpha=-3$, with six rather symmetric peaks roughly centered at
half-integral values of $\overline{B}$.  However, as for $S=1/2$, the low-$T$
limiting behavior is not reached until $\alpha=-10$.  For $S=5/2$, Fig. 3
shows that $C_V$  consists of  five  double peaks symmetrically centered about
$\overline{B}=s+1$ for $s=0,\ldots,4$, and nearly vanishes at those
points.

We note that this multiplicity of double peaks at low $T$ is also a
consequence of quantum level crossing.  At low $T$, the energy difference
between the two lowest states is given by Eq. (\ref{deltaE}).  At the exact
level crossing, these levels are degenerate, leading to an exponentially small
value of
$C_V$, [$\propto\exp(2\alpha)$]  Just away from these points, excitations
from the ground state to the first excited state can occur for
$k_BT\agt|J||s+1-|\overline{B}||$.  This should lead to a peak on each side of
the level crossing points, with a splitting $\Delta\overline{B}$ of the double peaks approximately
equal to  $1/|\alpha|$.  From the data in Figs. 2 and 3 for $|\alpha|=3,10$,
the splitting is  $\approx2.4/|\alpha|$, which is $\propto T$, but 2.4 times as
large as in the above crude estimate.

We remark that the $2S$ double peak
structures in $C_V$  shown  in Figs. 2 and 3 comprise a new quantum effect.
Although it may be  difficult to increase
$B$ to 40 T while holding $T$ fixed at  $\alt1$ K, it
might be easier to lower $T$ in a fixed, strong $B$.  From
Figs. 2 and 3, for $\overline{B}$ fixed at 0.8 or 1.2,  non-monotonic
$C_V(T)$ behavior upon lowered $T$ is predicted.  However, at $\overline{B}=1$, the behavior at low $T$ is
very different, as $C_V$ at $\alpha=-10$ is vanishingly small.  This striking
sensitivity to the precise value of $B$ is pictured in Fig. 4
for $S=5/2$. In Fig. 4, for $k_BT/|J|>1$, the
$T$, $B$ dependencies of $C_V$ are monotonic.    Aside from the rather
ordinary peak for $B=0$, the unusual behavior is illustrated by comparing the
curves for $\overline{B}=0.8, 1.0, 1.2$ for $k_BT/|J|<0.5$  The curves for   $\overline{B}=0.8,
1.2$ both have peaks at low $T$, but the curve for  the intermediate value
$\overline{B}=1.0$ does not.  This highly sensitive  dependence of $C_V$ upon
$T, B$ 
is a new quantum level crossing effect.

\section{Spin Dynamics}
We now evaluate the time autocorrelation function for both 
quantum and classical spin dimers for ${\bf B}\ne0$.  

\subsection{Quantum Spin Dynamics}

The time evolution of quantum spins is given by the commutator of the spin
operator with the Hamiltonian,
\begin{eqnarray}
i\dot{\bf S}_i(t)&=&[{\bf S}_i,{\cal H}],\\
{\bf S}_i&=&\exp(it{\cal H}){\bf S}_i(0)\exp(-it{\cal H}).
\end{eqnarray}
It is easiest in the quantum case just to keep the time dependence in this
form, and then to let ${\cal H}$ operate on the eigenstates in the matrix elements of
the correlation function.  The total spin operators ${\bf s}^2$ and $s_z$ are
independent of $t$.

\subsection{Classical Spin Dynamics}
The two classical spins ${\bf S}_i(t)$ each precess according to 
classical Heisenberg dynamics,
\begin{eqnarray}
\dot{\bf S}_i&=&J{\bf S}_i\times{\bf S}+\gamma{\bf S}_i\times{\bf B},\label{S1oft}\\
\noalign{\rm where $|{\bf S}_i(t)|=1$, for $i=1, 2$, leading to}\nonumber\\
\dot{\bf S}&=&\gamma{\bf S}\times{\bf B}.\label{Soft}
\end{eqnarray}
Hence, the total spin ${\bf S}$ precesses about ${\bf B}$, and the ${\bf
S}_i$ precess about both ${\bf B}$ and ${\bf S}$.
The solution to these equations  is given in the Appendix.

\section{Time Correlation Functions}

\subsection{Quantum Spins}

In this section, we evaluate the time correlation functions for quantum spins
with general spin value $S$. 
Here we calculate the time autocorrelation function for the quantum dimer.
For easy comparison with the classical results, we normalize the correlation
functions such that ${\rm Re}\>\>{\cal C}_{11}(0)=1$.\cite{MSL}   Then
${\cal C}_{11}(t)=\langle {\bf S}_1(t)\cdot{\bf S}_1(0)\rangle$, or
\begin{eqnarray}
{\cal C}_{11}(t)&=&\sum_{s=0}^{2S}\sum_{m=-s}^s\langle sm|e^{i{\cal H}t}{\bf
S}_1(0)e^{-i{\cal H}t}\cdot{\bf S}_1(0)\nonumber\\
& &\qquad\times e^{-\beta{\cal H}}|sm\rangle/D,\label{c11oft}\\
&=&\sum_{s,m}\sum_{s',m'}\sum_{\alpha=1}^3e^{it(E_{sm}-E_{s'm'})}e^{-\beta
E_{sm}}\nonumber\\
& &\qquad\times\Bigl|\langle sm|S_{1\alpha}|s'm'\rangle\Bigr|^2/D,\label{quantum}\\
-\beta E_{sm}&=&\alpha s(s+1)+bm,\\
D&=&S(S+1)Z,
\end{eqnarray}
where $Z$ is given by Eq. (\ref{pf}).  In Eq. (\ref{quantum}), the
expectation values are related to Clebsch-Gordon coefficients.

For quantum spins, the near-neighbor correlation function and the
autocorrelation function are related by the sum rule,
\begin{eqnarray}
\langle s_z^2\rangle+\langle{\bf s}^2-s_z^2\rangle\cos(\tilde{B}t)&=&
{1\over{2}}\langle{\bf s}(t)\cdot{\bf s}(0)+{\bf s}(0)\cdot{\bf s}(t)\rangle
\nonumber\\
&=&2S(S+1)\nonumber\\
& &\times{\rm Re}[{\cal C}_{12}(t)+{\cal C}_{11}(t)],\label{sumrule}
\end{eqnarray}
where 
 ${\cal C}_{12}(t)$ is normalized as is ${\cal C}_{11}(t)$. The
thermal averages $\langle{\bf s}^2\rangle$ and $\langle s_z^2\rangle$ can be written in the notation of
Eqs. (\ref{dos})-(\ref{bsofb}) as $\langle
s(s+1)\rangle$ and $\langle s(s+1)\rangle-\coth(b/2)\langle B_s(b)\rangle$, respectively.

\subsection{Classical Spins}

It is interesting to compare the quantum time correlation functions with those
obtained from classical spin dynamics.  For classical spin dimers, one solves the
equations of motion of the spins directly, taking the length of each spin
to be unity.  We first solve 
Eq. (\ref{Soft}) for ${\bf S}(t)$, and then  Eq. (\ref{S1oft}) to obtain
the ${\bf S}_i(t)$.  We then evaluate the classical autocorrelation
function ${\cal C}_{11}(t)=\langle {\bf S}_1(t)\cdot{\bf S}_1(0)\rangle$ by
averaging over the length $S$ ($0\le S\le2$) of the total spin,
the angle $\theta$ between  ${\bf S}$ and ${\bf B}$, and the  angle
$\phi_0$ describing the initial relative configuration of ${\bf S}_1$ and ${\bf S}_2$,
and requiring 
${\cal C}_{11}(0)=1$.  Since the
procedure is  analogous to that used for the  
four-spin ring,\cite{KL} the
results are given in the Appendix.  Here we only remark that the classical
correlation functions also must satisfy a sum rule,
\begin{eqnarray}
{1\over{6}}[1+2\cos(\tilde{B}t)]\langle S^2\rangle&=&{\cal C}_{11}(t)+{\cal C}_{12}(t).
\end{eqnarray}

\section{Fourier transforms}

\subsection{Quantum Spins}

The Fourier transform $\tilde{\cal C}_{11}(\omega)$ of the real part of the
autocorrelation function ${\cal C}_{11}(t)$ for the quantum dimer 
of spin-$S$ spins  is given by
\begin{eqnarray}
\tilde{\cal C}_{11}(\omega)&=&\sum_{i=0}^{6S+1}f_{S,i}[\delta(\omega-\omega_{Si})+\delta(\omega+\omega_{Si})].
\end{eqnarray}

The discrete mode frequencies $\omega_{S,i}$ and their amplitudes $f_{S,i}$
are given for $S=1/2$ and $S=5/2$ in Tables II and III in the Appendix.

\subsection{Classical Spins}

In order to compare the quantum and classical results, it is useful to evaluate the
Fourier transform  $\tilde{\cal
C}_{11}(\omega) = \int_{-\infty}^{\infty}dt \exp(i\omega t){\cal C}_{11}(t)$
of the classical autocorrelation function.
Letting $\tilde{\omega}=\omega/|J|$ and $\overline{B}=\tilde{B}/|J|$, we find for positive $\omega$  that
\begin{eqnarray}
\tilde{\cal
C}_{11}(\omega)&=&\delta(\tilde{\omega})C_{00}+
\delta(\tilde{\omega}-\overline{B})C_{01}+\delta\tilde{\cal C}_{11}(\tilde{\omega}).\label{cft}
\end{eqnarray}
Exact expressions for the discrete amplitudes $C_{00}$ and $C_{01}$, and for
the continuous part $\delta\tilde{\cal C}_{11}(\tilde{\omega})$ are given in the Appendix.
Then, in order to check the numerical evaluation of the above quantities, we employ the frequency sum rule,
$1=\int_0^{\infty}d\omega\tilde{\cal C}_{11}(\omega)/\pi$.

We note that $C_{00}$ also appears in the expression for $\tilde{\cal
C}_{11}(\omega)$ with $\omega\le0$.
 The coefficients $C_{00}$ and $C_{01}$ of the delta functions at
$\omega=0$ and $\tilde{\omega}=\overline{B}$, respectively, are functions of both
$\alpha$ and $\overline{B}$.  These
quantities arise from the long-time rms limit of ${\cal C}_{11}(t)$ and for
the long-time oscillatory behavior at the driving frequency,
respectively. This second delta function occurs at the resonant frequency in
 an EPR experiment.

\section{Numerical Results}
\subsection{Autocorrelation Function Spectra}

In Fig. 5, we plot the amplitudes $f_{5/2,i}$ of the delta-function modes of the
zero-field quantum $S=5/2$ AFM $\tilde{\cal C}_{11}(\omega)$ for positive 
$\omega/|J|$,
at the temperatures $\overline{T}=k_BT/[|J|S(S+1)]=0.1, 1, \infty$.
The amplitudes of the modes at these three ${\overline T}$ values are indicated by the symbols +, $\times$, *.
For comparison, we also show the results of the zero-field classical
calculation $\tilde{\cal C}_{11}(\omega)$.  For $\overline{B}=0$, both of the delta-function amplitudes
$C_{00}$ and $C_{01}$ appear at $\omega=0$, and their combined weight at these
 $\overline{T}$ values is
indicated by the  circle, triangle, and square, respectively.  In
addition, the continuous part $\delta\tilde{\cal C}_{11}(\omega)$ is plotted both
by scaling its amplitude by $1/[S(S+1)]^{1/2}$ and the frequency by
$[S(S+1)]^{1/2}$.  This combined scaling allows us to compare with the
quantum results, while  preserving the area under the curves.  For $S=5/2$, this scaling changes the maximum
$\tilde{\omega}=\omega/|J|$ from 2 to $\sqrt{35}\approx5.91$.  We note that
as $T\rightarrow\infty$, the quantum and classical delta functions at
$\omega=0$ agree exactly, both having the weight $\pi/2$.  Also as $T\rightarrow\infty$, the
classical $\delta\tilde{\cal C}_{11}(\omega)$ forms an envelope for the
amplitudes of the quantum delta functions,   except for a slight deviation
at the larger frequencies.  At $\overline{T}=1$, the agreement between the
quantum and classical results is also pretty good, although there are
deviations at nearly every quantum mode value.  These deviations are more
pronounced at $\overline{T}=0.1$, but the overall agreement is still rather
good.

\begin{figure}[t]
\includegraphics[width=0.45\textwidth]{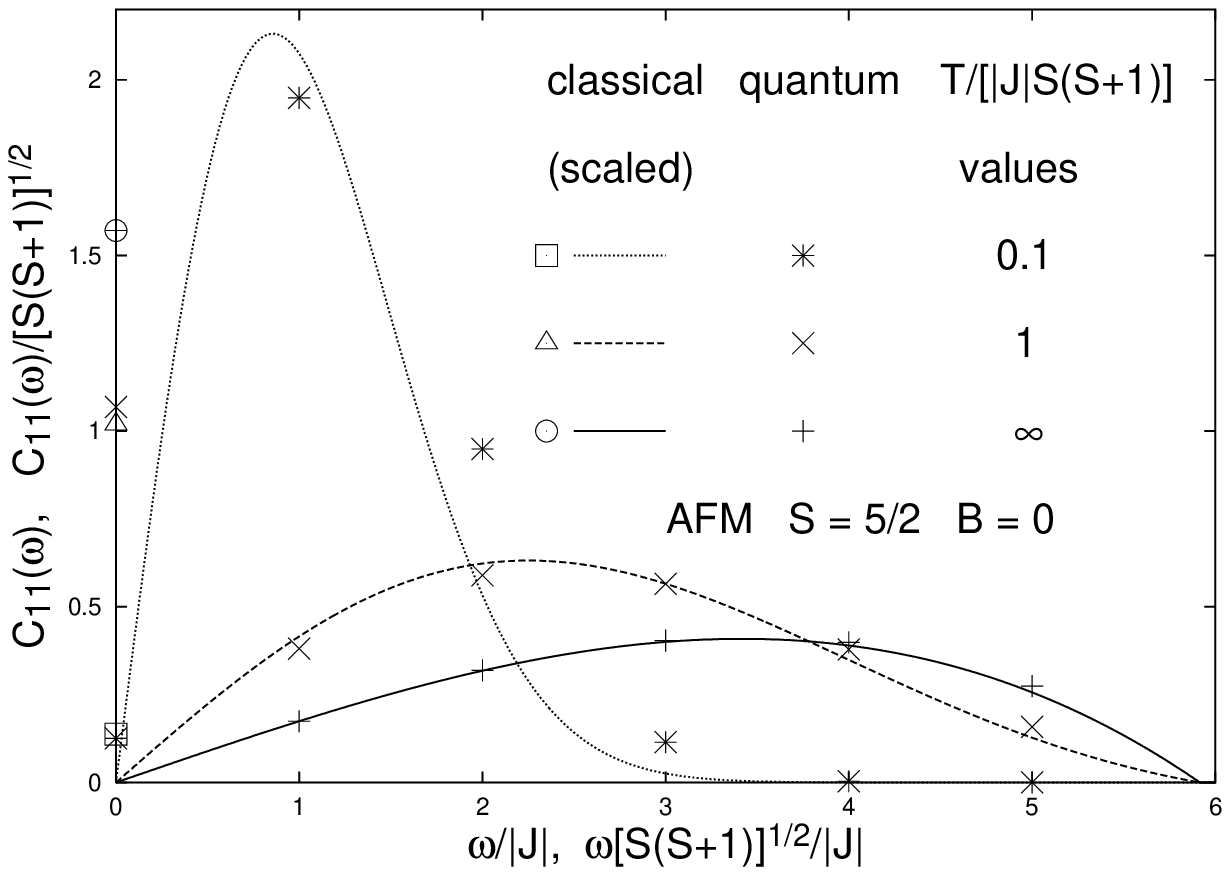}
\caption{Plots of the AFM quantum $S=5/2$ modes of $\tilde{\cal C}_{11}$ versus
$\omega/|J|$ at $\overline{B}=0$ at 
$\overline{T}=k_BT/[|J|S(S+1)]=0.1, 1, \infty$, and for the combined discrete classical modes at 
$\omega=0$.
The curves represent $\delta\tilde{\cal
C}_{11}(\omega)/[S(S+1)]^{1/2}$ versus $\omega[S(S+1)]^{1/2}/|J|$.}\label{fig5} 
\end{figure}

\begin{figure}[t]
\includegraphics[width=0.45\textwidth]{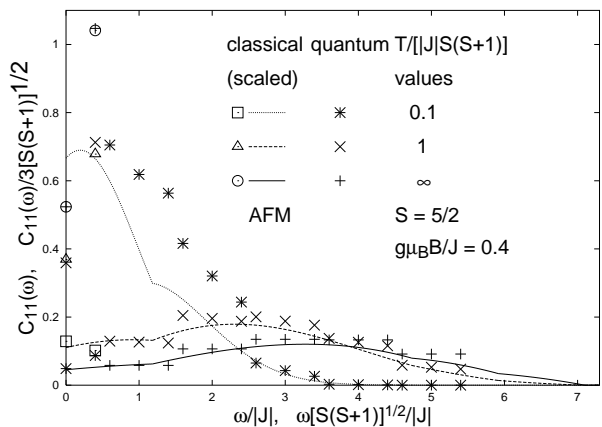}
\caption{Plots of the AFM quantum $S=5/2$ modes of $\tilde{\cal C}_{11}$ versus $\omega/|J|$ at $\overline{B}=0.4$ at 
$\overline{T}=0.1,1,\infty$, and for the discrete classical modes at $\omega/|J|=0,\overline{B}$.
The curves represent $\delta\tilde{\cal C}_{11}/3[S(S+1)]^{1/2}$
versus $\omega[S(S+1)]^{1/2}/|J|$.}\label{fig6} 
\end{figure}

In Fig. 6, we show the corresponding curves for the $S=5/2$ AFM dimer
at the  field  $\overline{B} = 0.4$.  For $\overline{B}\ne0$, each of the quantum zero-field modes splits into three
modes, and therefore to compare the continuous part $\delta\tilde{\cal
C}_{11}(\omega)$ with the split quantum modes, we divide the scaled continuous
part of $\tilde{\cal C}_{11}$ by 3, plotting $\delta\tilde{\cal C}_{11}/3[S(S+1)]^{1/2}$
versus $\omega[S(S+1)]^{1/2}/|J|$. As in Fig. 5, the classical delta functions
modes at $\tilde{\omega}=0,\overline{B}$ are not scaled, but in this case,
they are distinct.
For this rather strong field value, $\overline{B}=0.4$, the agreement
between the classical and quantum results is very good as
$T\rightarrow\infty$, and pretty good at the intermediate 
$\overline{T}=1$.  However, at $\overline{T}=0.1$,
these results can differ by more than a factor of two.
  We note that the classical curves are piecewise
continuous functions of $\omega$, with several discontinuities in slope
evident.  These discontinuities in slope generally appear at the the
frequencies corresponding to rather large jumps in the quantum mode amplitude  values. At $\overline{T}$ values lower than those pictured, the classical curves
develop into three distinct modes with finite widths. 

\begin{figure}[t]
\includegraphics[width=0.45\textwidth]{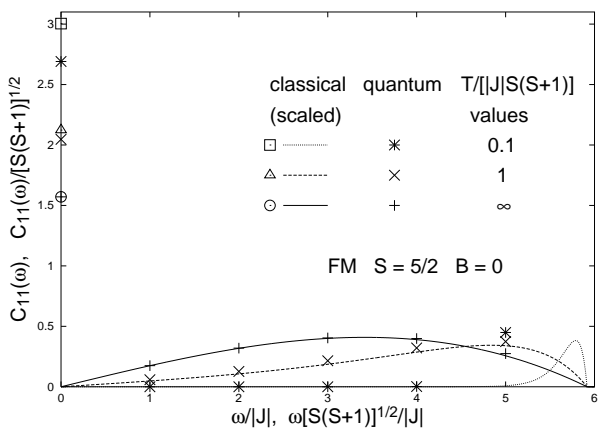}
\caption{Plots of the FM quantum $S=5/2$ modes of $\tilde{\cal C}_{11}$ versus
$\omega/|J|$ at $\overline{B}=0$ at 
$\overline{T}=0.1,1,\infty$, and for the combined discrete classical modes
 at 
$\omega=0$.
The  curves represent $\delta\tilde{\cal
C}_{11}(\omega)/[S(S+1)]^{1/2}$ versus $\omega[S(S+1)]^{1/2}/|J|$.}\label{fig7} 
\end{figure}

\begin{figure}[b]
\includegraphics[width=0.45\textwidth]{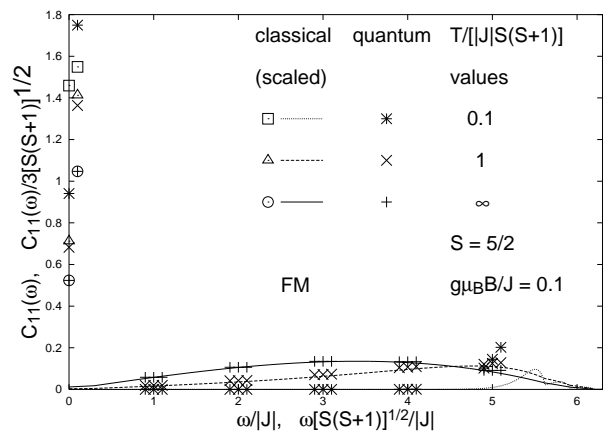}
\caption{Plots of the FM quantum $S=5/2$ modes of $\tilde{\cal C}_{11}$ 
versus $\omega/|J|$ at $\overline{B}=0.1$ at 
$\overline{T}=0.1, 1, \infty$, and for the discrete classical modes at
$\omega/|J|=0, \overline{B}$.
The  curves represent $\delta\tilde{\cal C}_{11}/3[S(S+1)]^{1/2}$
versus $\omega[S(S+1)]^{1/2}/|J|$.}\label{fig8} 
\end{figure}

In Figs. 7 and 8, we show the results for the FM $S=5/2$ case at
$\overline{B}=0, 0.1$, respectively.  Choosing $\overline{B}=0.1$ in Fig. 8
allows us to display all of the results clearly on the same figure.  Except
for the different $\overline{T}$ values in
Figs. 8 and 6, Figs. 7 and 8 correspond precisely to
Figs. 5 and 6, with the same scaling of the continuous part of the Fourier
transform of the classical correlation function, and the same symbols and
linestyles.    For
the FM case at $\overline{B}=0, 0.1$, Figs. 7 and 8 show that the agreement between the classical
and quantum modes is excellent as $T\rightarrow\infty$, pretty good at
$\overline{T}=1$, but only qualitative at
$\overline{T}=0.1$.  In this case, the classical delta function at $\omega=0$
differs significantly from the quantum one, and the continuous classical curve
cannot be scaled in this way at all.  As shown for the equivalent neighbor
model, the FM classical curves obey a different low-$T$ scaling relation, as
the low $T$ peak center approaches $NJ$ linearly in $T$ from below as
$T\rightarrow0$.\cite{KA}   In Fig. 8, the development of the continuous classical
curve into three low $T$ modes is evident, but the fit to the quantum case at
$\overline{T}=0.1$ is
not good at all.  Not only should the low-$T$ classical curves require a
different scaling relation, similar to those for $B=0$, the order of the relative heights of
the three largest modes at the largest frequencies pictured  for the classical
curves is inverted relative to that for the quantum case. 

\begin{figure}[t]
\includegraphics[width=0.45\textwidth]{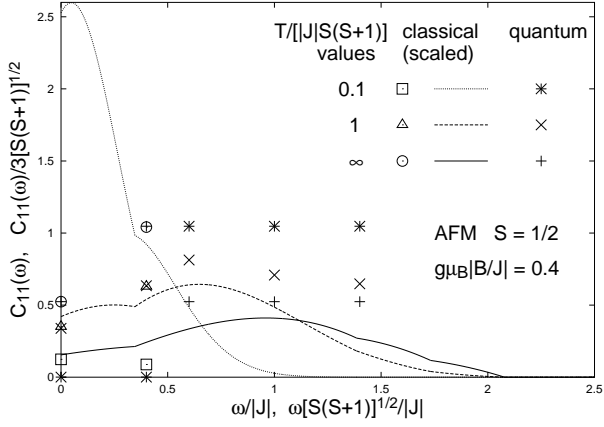}
\caption{Plots of the AFM quantum $S=1/2$ modes of $\tilde{\cal C}_{11}$ 
versus $\omega/|J|$ at $\overline{B}=0.4$ at 
$\overline{T}=0.1,1,\infty$, and for the discrete classical modes at $\omega/|J|=0,\overline{B}$.
The curves represent $\delta\tilde{\cal C}_{11}/3[S(S+1)]^{1/2}$
versus $\omega[S(S+1)]^{1/2}/|J|$.}\label{fig9} 
\end{figure}

\begin{figure}[b]
\includegraphics[width=0.45\textwidth]{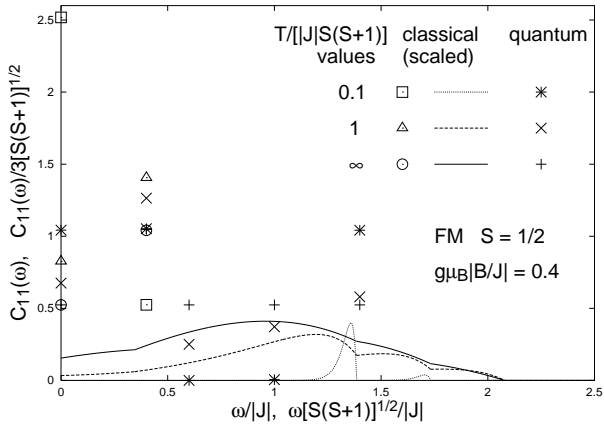}
\caption{Plots of the FM quantum $S=1/2$ modes of $\tilde{\cal C}_{11}$ versus
$\omega/|J|$ at $\overline{B}=0.4$ at $\overline{T}=0.1,1,\infty$, and for the discrete classical modes at $\omega/|J|=0,\overline{B}$.
The curves represent $\delta\tilde{\cal C}_{11}/3[S(S+1)]^{1/2}$
versus $\omega[S(S+1)]^{1/2}/|J|$.}\label{fig10} 
\end{figure}

We now consider the case $S=1/2$.  For the $B=0$ AFM and FM cases, there are two
quantum modes for $0\le\omega$, at $\omega/|J|=0, 1$.  
As $T\rightarrow\infty$, these
modes have equal intensity ($\pi/2$), for both FM and AFM cases.  For the AFM
case as $T$ decreases to 0, the amplitudes of
the modes at $\omega/|J|=0 (1)$  decrease (increase)  continuously to  0
($\pi$), respectively. For the FM case, however,  these mode amplitude values
increase (decrease)
with decreasing $T$ 
to   $2\pi/3$ ($\pi/3$), respectively,  at $T=0$. Somewhat surprisingly, for 
 the AFM case, the amplitude of the combined classical
delta functions at $\omega=0$ closely tracks that of the quantum $\omega=0$
mode.  For the FM case, however, this tracking is not so good.  However, the best that can be said for the continuous part is that as
$T\rightarrow\infty$, the second quantum mode nearly falls upon the classical curve
scaled as in Figs. 5 and 7, for both AFM and FM cases.  At $\overline{T}=k_BT/[|J|S(S+1)]=1, 0.1$, this agreement becomes
increasingly much worse, respectively.  For the FM case, the classical
treatment fails miserably as $T\rightarrow0$, as the amplitude of the
classical delta function mode at $\omega=0$ approaches $\pi$, and the
integrated intensity of the continuous classical mode peak vanishes as
$T\rightarrow0$.  Hence, the classical treatment does not describe the $S=1/2$
dimer at $B=0$ correctly, and the correct quantum treatment leads to just two
modes for $\omega\ge0$.

For $B\ne0$, however, there are five modes for $\omega\ge0$ with $S=1/2$, so
a classical treatment can approximate the quantum behavior a bit better than
for $B=0$.  In Figs. 9 and 10, the AFM and FM $S=1/2$ results for
$\overline{B}=0.4$ at $\overline{T}=0.1, 1, \infty$ are shown.  In both cases, the
two quantum modes for $\omega/|J|=0,\overline{B}=0.4$ are compared with the two classical delta
functions, and the remaining three quantum modes for $0\le\omega$ are compared
with the continuous classical curves.  In both cases, as $T\rightarrow\infty$, the five
quantum modes have the amplitudes $\pi/6, \pi/3, \pi/6, \pi/6, \pi/6$, respectively.
The classical results agree precisely with the first two, and form a
qualitative envelope similar to the remaining three quantum modes.  This
agreement is qualitatively preserved at $\overline{T}=1$.  However, as $T\rightarrow0$,
 the classical and quantum cases differ dramatically.   For the AFM case
pictured in Fig. 9, the five quantum
modes for $0\le\omega$
approach the  amplitudes 0, 0, $\pi/3,\pi/3,\pi/3$ as $T\rightarrow0$, respectively, and the continuous
classical curve develops a strong peak at $\omega=0$, which is qualitatively
different.  The qualitative behavior of the two discrete classical AFM modes
is not too bad, however.  For the FM case pictured in Fig. 10,  as
$T\rightarrow0$, the quantum FM modes approach the amplitudes
$\pi/3,\pi/3,0,0,\pi/3$, respectively. That is, the modes at $\omega=0,
|J+\tilde{B}|$ increase, the modes at $\omega=|J|, |J-\tilde{B}|$ decrease, and
the mode at $\omega=\tilde{B}$ has a non-monotonic $T$ dependence.  The
classical treatment preserves  these FM features only qualitatively, and
is inaccurate for $\overline{T}\le0.1$.

\subsection{Field dependencies of the quantum modes}

In Fig. 11, we plotted the amplitudes $f_{1/2,i}$ of the five FM quantum modes for $S=1/2$ at
 $\overline{T}=0.1$ versus $\overline{B}$.  The modes $\omega_{1/2,i}$
for $i=0,4$ correspond to $\omega=0$, $|\tilde{B}|$, $|J|$, $|J-\tilde{B}|$,
and $|J+\tilde{B}|$, respectively.  Note that at $B=0$, the $\omega_{1/2,i}$
for $i=0,1$ are degenerate but unequal in intensity, and the $\omega_{1/2,i}$
for $i=2,3,4$ are both degenerate and equal in intensity.  At each field, the sum
of the five intensities sums to $\pi$.  From this figure, it is evident that
the crossover from the low-field regime to the high field regime occurs at
the rather low field value, $\overline{B}\approx0.1$.  The high field
regime is clearly consistent with the mode amplitudes at $\overline{B}=0.4$
pictured in Fig. 10, with the amplitudes of the  modes at
$\omega/|J|= 0, 0.4, 0.6, 1, 1.4$ approximately equal to $\pi/3,
\pi/3,0,0,\pi/3$, respectively.

\begin{figure}[t]
\includegraphics[width=0.45\textwidth]{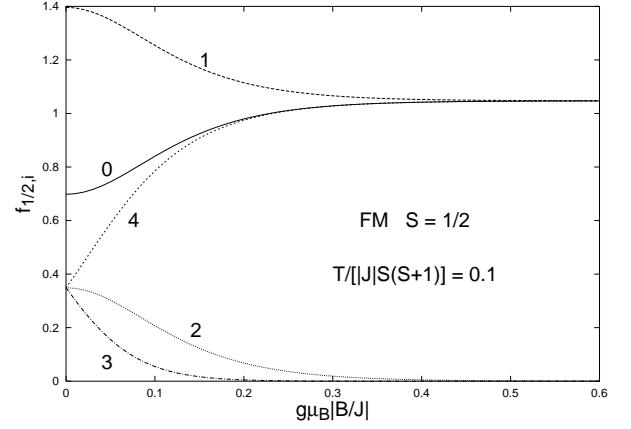}
\caption{Plots of the  FM quantum $S=1/2$ mode amplitudes $f_{1/2,i}$ versus
 $\overline{B}$ at $\overline{T}=0.1$  The numerical labels correspond to the $i$ values in
Table I of the Appendix.}\label{fig11} 
\end{figure}

\begin{figure}[b]
\includegraphics[width=0.45\textwidth]{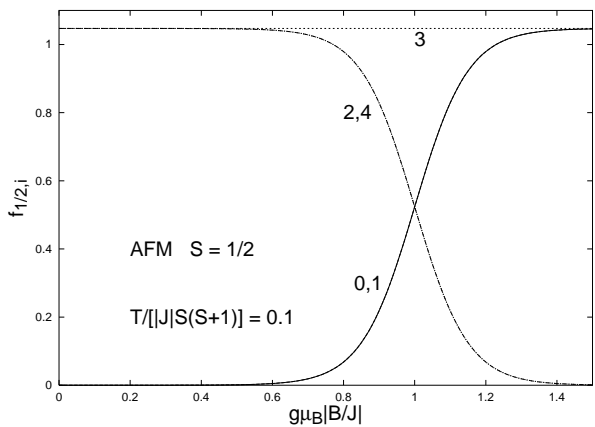}
\caption{Plots of the   AFM quantum $S=1/2$ mode amplitudes $f_{1/2,i}$ versus
$\overline{B}$ at 
$\overline{T}=0.1$.  The mode $i$ values correspond to Table I in the Appendix.}\label{fig12} 
\end{figure}

In Fig. 12, we  plotted  the amplitudes
$f_{1/2,i}$ of the five  AFM quantum modes for $S=1/2$ at the  
 $\overline{T}=0.1$.    From Fig. 12, the five modes are
difficult to discern clearly, due to the strong degeneracies involved.
Clearly, the crossover from the low-field to the high-field regime occurs at
$\overline{B}=1$.  In the low-field regime, the modes at
$\omega=0,\tilde{B}$ have nearly zero amplitudes, and the modes with
$\omega=|J|,|J\pm\tilde{B}|$ have nearly equal amplitude  $\pi/3$.  This
is the situation pictured for $\overline{B}=0.4$ in Fig. 9.  However, Fig. 12
indicates that dramatic changes in the mode amplitudes at low $T$ should take place as
$\overline{B}$ is increased from $\approx 0.8$ to $\approx 1.2$.  Over this field
range, the amplitudes of the modes at $\omega=|J|,|J+\tilde{B}|$ decrease
from nearly $\pi/3$ to nearly 0, and the amplitudes of the modes at
$\omega=0,\tilde{B}$ increase from nearly zero to nearly $\pi/3$.
Meanwhile, the
amplitude of the remaining mode at $\omega=|J|$ remains constant at $\pi/3$.
Although not pictured for brevity, at $\overline{T}=1$, all
five AFM modes for $S=1/2$  are clearly evident, and the crossover from the weak field regime
to the strong field regime occurs at $\overline{B}\approx1.5$.

\begin{figure}[t]
\includegraphics[width=0.45\textwidth]{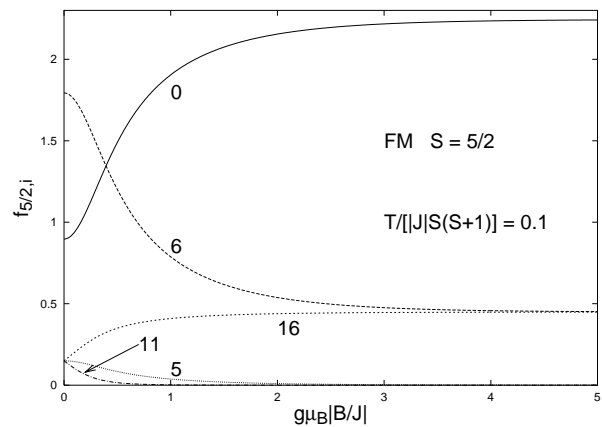}
\caption{Plots of  the dominant FM quantum $S=5/2$ mode amplitudes $f_{5/2,i}$ versus
 $\overline{B}$ at 
$\overline{T}=0.1$.  The mode $i$ values correspond to table II in the Appendix.}\label{fig13} 
\end{figure}

\begin{figure}[b]
\includegraphics[width=0.45\textwidth]{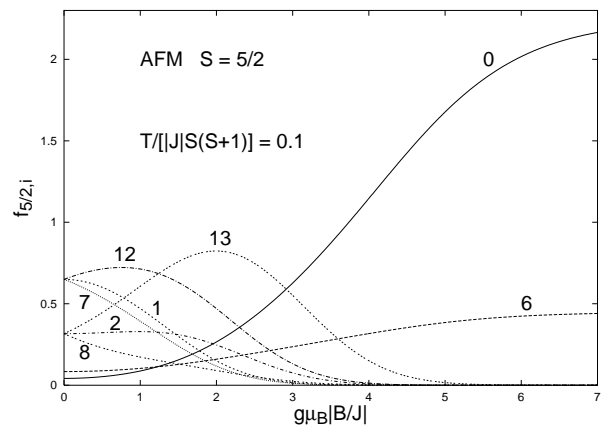}
\caption{Plots of the  dominant AFM quantum $S=5/2$ mode amplitudes
$f_{5/2,i}$ versus $\overline{B}$ at
$\overline{T}=0.1$.  The mode $i$ values correspond to table II in the Appendix.}\label{fig14} 
\end{figure}

The field dependencies of the most important modes for the FM and AFM $S=5/2$ dimers are
shown at the rather low temperature $\overline{T}=0.1$ in Figs. 13 and 14,
respectively.  In Fig. 13, the FM modes shown are for $\omega=0$, $\tilde{B}$,
$|5J\pm\tilde{B}|$, and $5|J|$.  The crossover from the weak to the strong
field limits occurs at about $\overline{B}\approx0.5$.  In the weak field limit, this corresponds to
the dominant modes at $\overline{B}=0.1$ shown in Fig. 8, for which the ranking
of the five strongest low-$T$ modes is at $\omega/|J|=0.1, 0, 5.1, 5, 4.9$,
respectively.  At high fields, $\overline{B}>2$, there are only three important
modes, at $\omega=0,\tilde{B}$, and $5J+\tilde{B}$, which have
amplitudes that approach $5\pi/7$, $\pi/7$, $\pi/7$, at $B\rightarrow\infty$ respectively. 
  
For the AFM case with $S=5/2$, the behavior of the relative mode amplitudes at
$\overline{T}=0.1$ is more complicated, as shown in Fig. 14.  There we 
plotted the field dependencies of the eight most important modes, which occur
at $\omega=0$, $|2J\pm\tilde{B}|$, $2|J|$, $|\tilde{B}|$,
$|J\pm\tilde{B}|$, $|J|$, respectively.  From Fig. 14, it is evident
 that there
are essentially three field regimes.  The low field regime occurs for $\overline{B}<0.5$,
the intermediate regime is for $1.5<\overline{B}<2.2$, and the high field
regime for $\overline{B}>3$, approximately.  An example of the low-field regime
at the same temperature was shown for $\overline{B}=0.4$ in Fig. 6. 
 In Figs. 6 and 14, the low-field rank-ordering of the six most important
modes are those at $\omega=|\tilde{B}+J|$, $|J|$, $|\tilde{B}-J|$,
$|\tilde{B}+2J|$, $2|J|$, and $|\tilde{B}-2J|$, where $J<0$. 
From   the intermediate regime pictured in Fig. 14, however, 
 the two most important
modes are those at $\omega=|\tilde{B}+2J|$ and $|\tilde{B}+J|$, with the six
other modes in the list contributing lesser, but comparable amounts.  Finally,
in the high-field regime, the two dominant modes are at $\omega=0$ and
$|\tilde{B}|$, respectively.  

\section{Discussion}
We remark that there have been some experiments on one or more of the Fe2
single molecule magnet dimers.  Le Gall {\it et al.} measured the four Fe2
dimers they made, and found that the zero-field magnetic susceptibility could
be fit with the Heisenberg model, with exchange constants ranging from 15 to
19 cm$^{-1}$ (22 to 27 K, which corresponds to $\overline{B}=1$ at $B\approx 16-22$ T). A slightly different fit was made by
Lascialfari {\it et al.}, obtaining $|J|\approx22K$ for the Fe2 dimer,
[Fe(OMe)(dpm)$_2$]$_2$. \cite{Fe2}  Those authors also refer to EPR measurements that
give rise to a zero-field splitting anisotropy of about 7 K in the first
excited state, which would complicate the analysis performed here.  Such a
zero-field splitting would still allow quantum level splitting, but the
magnetization steps and the specific heat double peaks would not all be
regularly spaced.  We remark that these  level crossing effects are purely
quantum in nature, as the analogous classical calculations do not give rise to
such effects, and hence are only approximate for
$|\alpha|\agt1$. Nevertheless, the quantum effects predicted for the
magnetization and specific heat ought to be observable with presently available
facilities, as one should be able to see one or two magnetization steps, and
one or two specific heat double peaks.

 In addition, inelastic neutron scattering would see
slightly different results from those presented here.  In this situation, the
classical envelope curves might be quite useful, as they can serve as a guide
to the behavior that might be expected with specific zero-field splitting
values.  More important, inelastic neutron scattering can be used to probe the
details of the magnetic interactions within an individual dimer.  Provided
that a single crystal of sufficient size for such studies can be obtained, one
would perform the experiments at the wave vectors specific to the crystal
structure.  More precisely, the unpolarized inelastic neutron magnetic
dynamical structure factor
$S({\bf q},\omega)$ for a single
crystal of equal-spin dimers  is given by \cite{Fe8,Cu2,white,fg}
\begin{eqnarray}
S({\bf
q},\omega)&=&\sum_{\alpha,\beta=1}^3(\delta_{\alpha\beta}-\hat{q}_{\alpha}\hat{q}_{\beta})\nonumber\\
& &\times\int_{-\infty}^{\infty}{{dt}\over{2\pi}}e^{i\omega t}\langle
Q^{\dag}_{\alpha}({\bf q},t)Q_{\beta}({\bf q},0)\rangle,\label{Sgeneral}
\end{eqnarray}
where the
$\hat{q}_{\alpha}=\sin\theta_q\cos\phi_q,\sin\theta_q\sin\phi_q,\cos\theta_q$
for $\alpha=1,2,3$, respectively, and $\theta_q$, $\phi_q$ are the angles
the scattering wave vector ${\bf q}$ makes with the spin quantization
coordinates, and $\langle\ldots\rangle$ represents a thermal average, as in
Eq. (\ref{c11oft}).  The dimer structure operator
\begin{eqnarray}
Q_{\alpha}({\bf q},t)&=&f({\bf q})[e^{i{\bf q}\cdot{\bf d}}S_{1\alpha}(t)+e^{-i{\bf
q}\cdot{\bf d}}S_{2\alpha}(t)],\label{Q}
\end{eqnarray}
where $2{\bf d}$ is the displacement vector between the dimer spins, and
$f({\bf q})$ is the atomic magnetic form factor.
 \cite{Cu2,fg}  The low $T$, $B=0$ transitions in $S=1/2$
Cu$^{2+}$ and Yb$^{3+}$ and in $S=3/2$ Cr$^{3+}$ dimer powders were 
treated previously.\cite{Cu2,Yb2,Cr2} For ${\bf B}=0$, the quantization axis
is  arbitrary, so
$\theta_q=\pi/2$ was chosen.\cite{Cu2} For ${\bf B}\ne0$, the quantization axis is
 parallel to
${\bf B}$, so $\theta_q$ and $\phi_q$ are the angles ${\bf q}$ makes with the
coordinates $\hat{\bf z}||{\bf B}$ and $\hat{\bf x}\times\hat{\bf
y}=\hat{\bf z}$. 
  For
scattering wavevectors ${\bf q}_c$ directed along the special angle  
\begin{eqnarray}
\theta_q^c&=&\sin^{-1}(2/3)^{1/2},
\end{eqnarray}
the components of each $\tilde{\cal C}_{ij}(\omega)$ are sampled equally,
and
\begin{eqnarray}
S({\bf q}_c,\omega)&=&{2\over{3\pi}}f^2({\bf q}_c)\bigl[\tilde{\cal
C}_{11}(\omega)+\tilde{\cal C}_{12}(\omega)\cos(2{\bf q}_c\cdot{\bf d})\bigr]
\nonumber\\
& &\\
&=&{2\over{3}}f^2({\bf q}_c)\Bigl[\tilde{\cal C}_{11}(\omega)[1-\cos(2{\bf
q}_c\cdot{\bf d})]/\pi+\nonumber\\
& &+{{\cos(2{\bf q}_c\cdot{\bf d})}\over{S(S+1)}}\Bigl(\langle
s_z^2\rangle\delta(\omega)+{1\over{2}}\langle {\bf s}^2-s_z^2\rangle\nonumber\\
& &\times[\delta(\omega-\tilde{B})+\delta(\omega+\tilde{B})]\Bigr)\Bigr],
\end{eqnarray}
where we have employed the Fourier transform of the sum rule in
Eq. (\ref{sumrule}). For powder samples, one can still use the special angle
technique with a field to obtain $\tilde{\cal C}_{11}(\omega)$, but since
the direction of ${\bf d}$ is  random, one obtains
\begin{eqnarray}
\overline{\cos(2{\bf q}_c\cdot{\bf d})}&=&{{\sin(2q_cd)}\over{2q_cd}},
\end{eqnarray}
where $\overline{\cdot\cdot\cdot}$ is a spatial average. \cite{fg}.

For the  general case of $\theta_q\ne\theta_q^c$, however, $S({\bf
q},\omega)$ cannot be written simply in terms of the $\tilde{\cal
C}_{ij}(\omega)$.   There are four factors
$h_i$ for $i=0,\ldots,3$, listed in the Appendix, that depend
upon ${\bf q}\cdot{\bf d}$ and $\sin\theta_q$.  We then find
\begin{eqnarray}
S({\bf q},\omega)&=&{{f^2({\bf
q})}\over{2\pi}}\sum_{i=1}^{6S+1}f_{S,i}h_{S,i}\nonumber\\
& &\times[\delta(\omega-\omega_{Si})+\delta(\omega+\omega_{Si})],
\label{Sqomega}
\end{eqnarray}
where  $h_{S,i}$ is the appropriate $h_i$ for the modes $\pm\omega_{Si}$, as
indicated for $S=1/2$ and 5/2 in the Appendix.  The factors $h_0$ and $h_1$ 
 are
$\propto\cos^2({\bf q}\cdot{\bf d})$,
and correspond respectively to the $\omega=0$ and $\pm\tilde{B}$ modes.   The
factors $h_2$ and $h_3$ are $\propto\sin^2({\bf q}\cdot{\bf d})$, and
 correspond
respectively to the modes at $\pm nJ$ and $\pm|nJ\pm\tilde{B}|$, for
$n=1,\ldots,2S+1$. Since $h_0$ and $h_2$ are also 
 $\propto\sin^2\theta_q$, whereas $h_1$ and $h_3$ are 
 $\propto(1-{1\over{2}}\sin^2\theta_q)$,  the experimenter can fine tune the 
 single crystal data by rotating ${\bf
B}$ and ${\bf q}$ relative
to ${\bf d}$. 

Neutron powder data on the deuterated $S=1/2$ dimer single molecule magnet 
VODPO$_4\cdot{1\over{2}}$D$_2$O (V2) were taken, resulting in a fit to the
AFM Heisenberg model of $|J|=7.81(4)$ meV, \cite{V2neutron} close to the value 7.6 meV found in
the susceptibility fit. \cite{Johnson}  This corresponds to $\overline{B}=1$ at
$B\approx 66$ T, which is too large for thermodynamic studies.
However, inelastic neutron scattering at 0.1-0.2$\overline{B}$ ought to be
possible for this material.  Single crystal data could be particularly
interesting.  For the Fe2 dimers, inelastic neutron scattering in a field
of $\overline{B}\alt0.4-0.5$ should be possible, which would not see any level
crossing effects, but could prove interesting, as indicated in Fig. 6.

\section{Acknowledgments}
We thank M. Ameduri and S. E.
 Nagler for useful discussions.  
DE gratefully acknowledges support from Project Number SFB463 with the
MPI-PKS, the MPI-CPfS, and  IFW, and the TU Dresden.

\section{Appendix}

\subsection{Classical Time Correlation Function}

From the classical equations of motion, 
\begin{equation}
{\bf S}(t)=\hat{\bf B}S_{||}+S_{\perp}[\hat{\bf x}\cos(\tilde{B}t)-\hat{\bf
y}\sin(\tilde{B}t)],
\end{equation}
where $\tilde{B}=\gamma B$, $\hat{\bf B}$ is a unit vector parallel to ${\bf B}$, and $\hat{\bf x}$,
and $\hat{\bf y}$ are orthogonal unit vectors satisfying $\hat{\bf
x}\times\hat{\bf y}=\hat{\bf B}$.
Since $S_{||}={\bf S}\cdot\hat{\bf B}$, we have
$S_{||}=S\cos\theta$ and
$S_{\perp}=S\sin\theta$, and hence
$S^2=S_{||}^2+S_{\perp}^2$.
 ${\bf S}_1(t)$ is then found to be
\begin{eqnarray}
S_{1,||}(t)&=&S_{1,||}+S_{1,\perp}\cos(JSt-\phi_0),\\
S_{1\pm}(t)&=&e^{\pm i\gamma Bt}\Bigl[{{S_{1,||}S_{\perp}}\over{S_{||}}}+{{
S_{1,\perp}S_{\perp}}\over{2}}\times\nonumber\\
& &\times\Bigl({{\exp[\mp i(JSt-\phi_0)]}\over{S_{||}-S}}+{{\exp[\pm
i(JSt-\phi_0)]}\over{S_{||}+S}}\Bigr)\Bigr],\nonumber\\
\end{eqnarray}
where $S_{1\pm}=S_{1x}\pm iS_{1y}$, and $\phi_0$ fixes the initial
relative configuration of the two spins. 
Since ${\bf S}_2={\bf S}-{\bf S}_1$, we have
$S_{1,||}=S_{2,||}=S_{||}/2$ and
$S_{1,\perp}=-S_{2,\perp}=(S_{\perp}/S)[1-S^2/4]^{1/2}$.
After averaging over $\phi_0$,
\begin{eqnarray}
{\cal C}_{11}(t)&=&{1\over{4}}\langle S_{||}^2+2{{S_{\perp}^2}\over{S^2}}(1-S^2/4)\cos(SJt)\nonumber\\
& &+S_{\perp}^2\cos(\tilde{B}t)+{{(1-S^2/4)}\over{S^2}}\Bigl((S+S_{||})^2\nonumber\\
& &\qquad\times\cos[(JS+\tilde{B})t]+
(S-S_{||})^2\nonumber\\
& &\qquad\times\cos[(JS-\tilde{B})t]\Bigr)\rangle.\label{phiaverage}
\end{eqnarray}
Replacing $S$ with $s$ for elegance, these classical averages are evaluated from 
\begin{eqnarray}
\langle\ldots\rangle&=&Z^{-1}\int_0^2sds\int_0^{\pi}{{\sin\theta
d\theta}\over{2}}e^{\alpha s^2+bs\cos\theta}\ldots,\\
Z&=&\int_0^2sds\int_0^{\pi}{{\sin\theta
d\theta}\over{2}}e^{\alpha s^2+bs\cos\theta}.
\end{eqnarray}
The integrals over $\theta$ can then  be written in terms of  $F_0(x)=\sinh(x)/x$ and its first and second derivatives,
$F_1(x)=F_0'(x)=[\cosh(x)-\sinh(x)/x]/x$
and $F_2(x)=F_0''(x)=F_0(x)-2F_1(x)/x$, respectively.  We note that
$F_1(x)/F_0(x)=L(x)=\coth(x)-1/x$ is the Langevin function.
We find

\begin{eqnarray}
{\cal C}_{11}(t)&=&{1\over{8Z}}\int_0^2sdse^{\alpha
s^2}\Bigl[2s^2G_1(bs,\tilde{B}t)\nonumber\\
& &+(4-s^2)\Bigl(\cos(sJt)G_2(bs,\tilde{B}t)\nonumber\\
& &+G_3(bs,\tilde{B}t)\sin(sJt)\Bigr)\Bigr],\label{C11oft}\\
G_1(x,y)&=&F_2(x)+[F_0(x)-F_2(x)]\cos(y),\\
G_2(x,y)&=&F_0(x)-F_2(x)+[F_0(x)+F_2(x)]\cos(y),\nonumber\\
& &\\
G_3(x,y)&=&2F_1(x)\sin(y).
\end{eqnarray}

As $T\rightarrow\infty$.  
we set  $t^*=|J|t$, and
obtain,
\begin{eqnarray}
\lim_{T\rightarrow\infty}{\cal C}_{11}(t)
&=&{{[1+2\cos(\tilde{B}t)]}\over{6}}f(t^*),\\
f(t^*)&=&1-{{[1+2\cos(2t^*)]}\over{t^{*2}}}\nonumber\\
& &+{{3\sin(2t^*)}
\over{t^{*3}}}-{{3[1-\cos(2t^*)]}\over{2t^{*4}}}.\label{foft}
\end{eqnarray}
We note that $f(t^*)$  was obtained previously for the zero field
case.\cite{Mueller}  We also have
$\lim_{T\rightarrow\infty}{\cal C}_{12}(t)=1-\lim_{T\rightarrow\infty}{\cal
C}_{11}(t)$.
These forms clearly satisfy the requirement $\lim_{T\rightarrow\infty}{\cal C}_{11}(0)=1$.

\subsection{Quantum Frequency Spectrum}

The Fourier transform $\tilde{\cal C}_{11}(\omega)$ of the real part of the
autocorrelation function ${\cal C}_{11}(t)$ for the quantum dimer 
of spin-$S$ spins  is given by
\begin{eqnarray}
\tilde{{\cal C}}_{11}(\omega)&=&\sum_{i=0}^{6S+1}f_{S,i}[\delta(\omega-\omega_{Si})+\delta(\omega+\omega_{Si})],\\
f_{S,i}&=&\pi a_{S,i}/{\cal C}^S(b,\alpha).
\end{eqnarray}
For both $S=1/2$ and $S=5/2$, ${\cal C}^S(b,\alpha)$ is given by
\begin{eqnarray}
{\cal
C}^S(b,\alpha)&=&4S(S+1)e^{2S[b-(2S+1)\alpha]}Z.
\end{eqnarray}
 We note that
$D=S(S+1)Z$ and that ${\cal C}^S(0,0)=4S(S+1)(2S+1)^2$.

The factors $h_i$ that weight the modes in 
$S({\bf q},\omega)$ given by Eq. (\ref{Sqomega}) can be derived from 
Eqs. (\ref{Sgeneral}) and (\ref{Q}). In Eq. (\ref{Sgeneral}), the off-diagonal
terms in $S({\bf q},\omega)$ with $\alpha,\beta=1,2$ sum to zero, and the remaining off-diagonal
terms all vanish. Hence, we only 
require
the matrix elements
\begin{eqnarray}
{\cal
M}_{sm,\alpha}^{s'm'}&=&(1-\hat{q}_{\alpha}^2)\Bigl|\langle 
sm|Q_{\alpha}^{\dag}({\bf q},0)|s'm'\rangle\Bigr|^2/f^2({\bf q}).
\end{eqnarray} 
From Eq. (\ref{m}) and $s_{\pm}|sm\rangle=A^{\pm}_{sm}|s,m\pm1\rangle$,
where $A^{\pm}_{sm}=[s(s+1)-m(m\pm1)]^{1/2}$, we write \cite{CS}
\begin{eqnarray}
\langle
sm|S_{nz}|s'm'\rangle&=&\delta_{m',m}[m\delta_{s',s}/2+(-1)^{n-1}\nonumber\\
& &\times(B_{sm}\delta_{s',s+1}+C_{sm}\delta_{s',s-1})],\label{s1z}\\
\langle
sm|S_{n\pm}|s'm'\rangle&=&\delta_{m',m\pm1}[A_{sm}^{\pm}\delta_{s',s}/2+(-1)^{n-1}\nonumber\\
& &\times(B_{sm}^{\pm}\delta_{s',s+1}+C_{sm}^{\pm}\delta_{s',s-1})],\label{s1pm}
\end{eqnarray}
for $n=1,2$, respectively. We note that $s_z=S_{1z}+S_{2z}$, $s_{\pm}=S_{1\pm}+S_{2\pm}$.  
We  find
\begin{eqnarray}
{\cal
M}_{sm,\alpha}^{s'm'}&=&\delta_{m',m}\bigl[\delta_{s',s}m^2h_0\nonumber\\
& &+4h_2\bigl(B_{sm}^2
\delta_{s',s+1}+C_{sm}^2\delta_{s',s-1}\bigr)\bigr]\nonumber\\
& &+\delta_{m',m\pm1}\Bigl[h_1\bigl(A_{sm}^{\pm}\bigr)^2\delta_{s',s}/2+2h_3
\nonumber\\
& &\times\Bigl(\bigl(B_{sm}^{\pm}\bigr)^2\delta_{s',s+1}+\bigl(C_{sm}^{\pm}
\bigr)^2\delta_{s',s-1}\Bigr)\Bigr],
\end{eqnarray}
where the
$h_i$  and the changes $\Delta s=s'-s$ and $\Delta m=m'-m$ in
the matrix elements for which they occur are listed in Table I. For arbitrary
$S$, each of the $h_{Si}$ is equal to one of the four $h_i$ listed in Table I.
For $S=1/2$ and 5/2, the appropriate choices of the $h_{Si}$ are listed in
Tables II and III, respectively.  We remark that by setting $f({\bf q})=1$ and each
of the $h_i=1$, $S({\bf q},\omega)\rightarrow\tilde{C}_{11}(\omega)/(2\pi)$.  

\begin{table}[t]
\begin{tabular}{c|c|c|c}
\hline
$i$&$\Delta s$&$\Delta m$&$h_i$\\
\hline
0&0&0&$\sin^2\theta_q\cos^2({\bf q}\cdot{\bf d})$\\
1&$0$&$\pm1$&$(1-{1\over{2}}\sin^2\theta_q)\cos^2({\bf q}\cdot{\bf d})$\\
2&$\pm1$&0&$\sin^2\theta_q\sin^2({\bf q}\cdot{\bf d})$\\
3&$\pm1$&$\pm1$&$(1-{1\over{2}}\sin^2\theta_q)\sin^2({\bf q}\cdot{\bf d})$\\
\hline
\end{tabular}
\caption{Factors $h_i$ that appear in $S({\bf q},\omega)$, Eq. (31), and their
associated transition quantum number changes.}\label{tab1}
\end{table}

\subsubsection{Quantum Frequencies for $S=1/2$}

For $S=1/2$, we have
\begin{eqnarray}
{\cal C}^{1/2}(b,\alpha)&=& 3(1+e^b+e^{2b}+e^{b-2\alpha}),
\end{eqnarray}

The mode frequencies $\omega_{1/2,i}$, their relative amplitudes  $a_{1/2,i}$,
and the factors $h_{1/2,i}$ appearing in $S({\bf q},\omega)$ are given in Table II.
\begin{table}[t]
\begin{tabular}{c|c|c|c}
\hline
$i$&$\omega_{1/2,i}$&$h_{1/2,i}$&$a_{1/2,i}$\\
\hline
0&0&$h_0$&$1+e^{2b}$\\
1&$|\tilde{B}|$&$h_1$&$(1+e^b)^2$\\
2&$|J|$&$h_2$&$e^b(1+e^{-2\alpha})$\\
3&$|J-\tilde{B}|$&$h_3$&$e^b(e^b+e^{-2\alpha})$\\
4&$|J+\tilde{B}|$&$h_3$&$1+e^{b-2\alpha}$\\
\hline
\end{tabular}
\caption{Frequency $\omega_{S,i}$
and weighting factor $h_{S,i}$ spectra of the $S=1/2$ Heisenberg  dimer in a magnetic
field.}\label{tab2}
\end{table}

\subsubsection{Quantum Frequencies for $S=5/2$}

For simplicity, we set
\begin{eqnarray}
A_s(b)&=&{{(2s+1)\sinh[(2s+1)b/2]}\over{\sinh(b/2)}},\\
X_s(b)&=&{{e^{-b/2}}\over{4\sinh^3(b/2)}}\Bigl(e^{sb}\sinh(sb)\nonumber\\
& &-s\sinh(b)-4s^2\sinh^2(b/2)\Bigr),\\
Y_s(b)&=&X_s(-b)={{e^{b/2}}\over{4\sinh^3(b/2)}}\Bigl(e^{-sb}\sinh(sb)\nonumber\\
& &-s\sinh(b)+4s^2\sinh^2(b/2)\Bigr)
\end{eqnarray}
We note that $X_1(b)=Y_1(b)=1$ and that $X_s(0)=Y_s(0)=s(4s^2-1)/3$.
The mode frequencies $\omega_{5/2,i}$, their relative amplitudes  $a_{5/2,i}$,
and the $h_{5/2,i}$ factors are given in Table III.
 In evaluating the coefficients $a_{5/2,i}$ at $\alpha=b=0$, it is useful to
employ the relations 
\begin{eqnarray}
\sum_{n=1}^{2S}\sum_{s=1}^ns^2&=&{{S(S+1)(2S+1)^2}\over{3}},\\
\sum_{s=0}^n(2s+1)^2&=&{{(n+1)(2n+1)(2n+3)}\over{3}}
\end{eqnarray}

\begin{table}[t]
\begin{tabular}{c|c|c|l}
\hline
$i$&$\omega_{5/2,i}$&$h_{5/2,i}$&$a_{5/2,i}$\\
\hline
0&0&$h_0$&$2e^{5(b-6\alpha)}\sum_{n=1}^{2S}e^{\alpha n(n+1)}\times$\\
& & &$\times\sum_{s=1}^{n}s^2\cosh(sb)$\\
1&$|J|$&$h_2$&${{35}\over{3}}(1+e^{-2\alpha})e^{5b-28\alpha}$\\
2&$2|J|$&$h_2$&${{32}\over{15}}e^{5b-24\alpha}(1+e^{-4\alpha})\sum_{s=0}^1A_s(b)$\\
3&$3|J|$&$h_2$&${{27}\over{35}}e^{5b-18\alpha}(1+e^{-6\alpha})\sum_{s=0}^2A_s(b)$\\
4&$4|J|$&$h_2$&${{140}\over{441}}e^{5b-10\alpha}(1+e^{-8\alpha})\sum_{s=0}^3A_s(b)$\\
5&$5|J|$&$h_2$&${1\over{9}}(1+e^{-10\alpha})e^{5b}\sum_{s=0}^4A_s(b)$\\
6&$|\tilde{B}|$&$h_1$&$2\coth(b/2)e^{5(b-6\alpha)}\sum_{n=1}^{2S}e^{\alpha n(n+1)}\times$\\
& & &$\times\sum_{s=1}^{n}s\sinh(sb)$\\
7&$|\tilde{B}+J|$&$h_3$&${{35}\over{3}}(1+e^{b-2\alpha})e^{4b-28\alpha}X_1(b)$\\
8&$|\tilde{B}+2J|$&$h_3$&${{32}\over{15}}e^{3b-24\alpha}(1+e^{b-4\alpha})X_2(b)$\\
9&$|\tilde{B}+3J|$&$h_3$&${{27}\over{35}}e^{2b-18\alpha}(1+e^{b-6\alpha})X_3(b)$\\
10&$|\tilde{B}+4J|$&$h_3$&${{140}\over{441}}e^{b-10\alpha}(1+e^{b-8\alpha})X_4(b)$\\
11&$|\tilde{B}+5J|$&$h_3$&${1\over{9}}(1+e^{b-10\alpha})X_5(b)$\\
12&$|\tilde{B}-J|$&$h_3$&${{35}\over{3}}e^{5b-28\alpha}(e^{b}+e^{-2\alpha})Y_1(b)$\\
13&$|\tilde{B}-2J|$&$h_3$&${{32}\over{15}}e^{6b-24\alpha}(e^b+e^{-4\alpha})Y_2(b)$\\
14&$|\tilde{B}-3J|$&$h_3$&${{27}\over{35}}e^{7b-18\alpha}(e^b+e^{-6\alpha})Y_3(b)$\\
15&$|\tilde{B}-4J|$&$h_3$&${{140}\over{441}}e^{8b-10\alpha}(e^b+e^{-8\alpha})Y_4(b)$\\
16&$|\tilde{B}-5J|$&$h_3$&${1\over{9}}e^{9b}(e^b+e^{-10\alpha})Y_5(b)$.\vspace{2pt}\\
\hline
\end{tabular}
\caption{Frequency $\omega_{S,i}$ and weighting factor $h_{S,i}$ spectra of the $S=5/2$ Heisenberg  dimer in a magnetic field.}\label{tab3}
\end{table}

\subsection{Classical Frequency Spectrum}

From Eq. (\ref{cft}) and letting $\tilde{\omega}=\omega/J$ and $\overline{B}=\tilde{B}/J$, the classical spin Fourier transform $\tilde{\cal
C}_{11}(\omega)$ has the following
discrete and continuous contributions, 

\begin{eqnarray}
\tilde{\cal
C}_{11}(\omega)&=&\delta(\tilde{\omega})C_{00}+\delta(\tilde{\omega}-
\overline{B})C_{01}+\delta\tilde{\cal C}_{11}(\tilde{\omega}),\\
\delta\tilde{\cal C}_{11}(\tilde{\omega})&= &
\sum_{i=1}^4C_i(\tilde{\omega}),\\
C_{00}&=&{{\pi}\over{4Z|J|}}\int_0^2s^3dse^{\alpha
s^2}F_2(bs),\\
C_{01}&=&{{\pi}\over{4Z|J|}}\int_0^2s^3dse^{\alpha
s^2}[F_0(bs)-F_2(bs)],\\
C_1(\tilde{\omega})&=&{{\pi}\over{4Z|J|}}\Theta(2-\tilde{\omega})2\tilde{\omega}(1-\tilde{\omega}^2/4)e^{\alpha \tilde{\omega}^2}\times\nonumber\\
& &\times[F_0(b\tilde{\omega})
-F_2(b\tilde{\omega})],\\
C_2(\tilde{\omega})&=&{{\pi}\over{4Z|J|}}\Theta(2-\overline{B}-\tilde{\omega})
(\tilde{\omega}+\overline{B})[1-(\tilde{\omega}+\overline{B})^2/4]\times\nonumber\\
& &\times e^{\alpha(\tilde{\omega}+\overline{B})^2}\Bigl(F_0[b(\tilde{\omega}+\overline{B})]+\nonumber\\
& &
F_2[b(\tilde{\omega}+\overline{B})]+2F_1[|b|(\tilde{\omega}+\overline{B})]\Bigr),\\
C_3(\tilde{\omega})&=&{{\pi}\over{4Z|J|}}\Theta(\overline{B}-\tilde{\omega})\Theta(\tilde{\omega}+2-\overline{B})(\overline{B}-\tilde{\omega})
\times\nonumber\\
& &\times [1-(\tilde{\omega}-\overline{B})^2/4]e^{\alpha(\overline{B}-\tilde{\omega})^2}\Bigl(F_0[b(\overline{B}-\tilde{\omega})]\nonumber\\
& &+F_2[b(\overline{B}-\tilde{\omega})]+2F_1[|b|(\overline{B}-\tilde{\omega})]
\Bigr),\\
C_4(\tilde{\omega})&=&{{\pi}\over{4Z|J|}}\Theta(\tilde{\omega}-\overline{B})\Theta(2+\overline{B}-\tilde{\omega})(\tilde{\omega}-\overline{B})\times\nonumber\\
& &\times e^{\alpha(\tilde{\omega}-\overline{B})^2}\Bigl(F_0[b(\tilde{\omega}-\overline{B})]
\nonumber\\
& &+F_2[b(\tilde{\omega}-\overline{B})]-2F_1[|b|(\tilde{\omega}-\overline{B})]
\Bigr).
\end{eqnarray}

\subsection{Low Temperature Classical Modes}

We now investigate the low $T$ behavior of the various contributions
$C_i({\omega})$ to the classical $\tilde{\cal C}_{11}(\omega)$. We follow the 
procedure  used for the
isosceles triangle and equivalent neighbor models in zero field. \cite{AK,KA}
 
The FM modal spectrum as $T\rightarrow0$ is 
given by
\begin{eqnarray}
\Omega_1(\overline{B})&=&|\overline{B}-2|,\\
\Omega_2(\overline{B})&=&2,\\
\Omega_3(\overline{B})&=&\overline{B}+2.
\end{eqnarray}
We note that $C_1$ leads to $\Omega_2$,  $C_2$ and $C_3$ combine  to create
$\Omega_1$, and $C_4$ leads to $\Omega_3$.  

  As $T\rightarrow0$, the AFM mode frequencies satisfy
\begin{eqnarray}
\Omega_1(\overline{B})&=&0\Theta(2-\overline{B})+(\overline{B}-2)\Theta(\overline{B}-2),\label{zero}\\
\Omega_2(\overline{B})&=&\overline{B}\Theta(2-\overline{B})+2\Theta(\overline{B}-2),\\
\Omega_3(\overline{B})&=&2\overline{B}\Theta(2-\overline{B})+(\overline{B}+2)\Theta(\overline{B}-2).
\end{eqnarray}
In $\Omega_1(\overline{B})$, the 0 indicates that the maximum of the mode is at
$\tilde{\omega}=0$, the same position as for $C_{00}$.

\end{document}